\documentstyle[aps,twocolumn,epsf,amssymb,floats]{revtex}

\newcommand{\PostScript}[3]{
  \vspace{#1cm}
\begin{center}
  \epsfysize=#2cm \leavevmode \epsfbox{#3}
\par
\end{center}
}

\begin{document}
\draft \preprint{preprint number}

\title{Wavepacket-current $\longleftrightarrow$ supercurrent
conversion in closed  {\em N/S/N} circuits}

\author{Arne Jacobs and Reiner K\"ummel}
\address{Institut f\"ur Theoretische Physik, Universit\"at W\"urzburg,
D-97074 W\"urzburg, Germany}

\date{\today} \maketitle
\begin{abstract}
The rise and decay in space and time  of a supercurrent in a superconductor
between two normal current leads connected to a reservoir is computed with
the help of quasiparticle wavepackets which suffer Andreev scattering in the
interfaces between the normal current leads and the superconductor. Stills
from a computer movie visualize electron $\leftrightarrow$ hole  scattering
and the creation and destruction of the supercurrent.

\end{abstract}


\narrowtext \flushbottom

Andreev scattering (AS) of electrons into holes and vice versa by spatial
variations of the superconducting pair potential \cite{and64}, in competition
and cooperation with conventional scattering,
determines the electronic structure and transport properties of
inhomogeneous superconductors. The Tomasch effect in tunnel junctions
\cite{mcm66}, Josephson currents \cite{kul70,ish70,bar72,gun94,kro99,jac99}, 
excess currents, and subharmonic gap structures \cite{blo82,oct83,kue90,gun94a} 
in superconducting ($S$)-normal
conducting ($N$)-superconducting junctions, as well as the transfer of half
of the Magnus force to the core electrons of a moving vortex line
\cite{hof} are due to AS. There is AS in He$^3$, too \cite{sch89}.
A wealth of AS phenomena has been discussed recently in \cite{bag99}
and \cite{oht99}.

While the conversion of a normal current into a supercurrent by electron
$\rightarrow$ hole scattering in the interface between an $N$ and an $S$
region of semi-infinite lengths has been described before
\cite{kue69,mat78,blo82}, it is the purpose of this paper to analyze the
normal-current $\longleftrightarrow$ supercurrent conversion processes in a
superconducting layer of finite length $L_z$ between two normal current
leads. These normal leads are connected to a reservoir (``battery'') which
acts as the current source in the closed circuit. The extensions of the
$N$ and $S$ regions in $x$- and $y$-directions are $L_x$ and $L_y$. The
metal-vacuum boundaries are treated as rigid walls. By varying $L_x$ and
$L_y$ one can vary the dimensionality of the system. The much investigated
quasi-two-dimensional (Q2D) superconducting/semiconducting heterojunctions
\cite{nit94,bas98} 
and superconducting quantum dots in Q2D channels
are systems for which our analysis may prove useful.

AS and the associated formation and destruction of Cooper pairs and
supercurrents can be calculated from the Time-dependent Bogoliubov-de
Gennes Equations (TdBdGE) \cite{kue69,mat78,kue99,blo82}. They describe the
evolution of the spinor quasiparticle (q.p.) wavefunction
with the electron component
$u_{n}({\bf r},t)$ and the hole component $v_{n}({\bf r},t)$ under the
influence of scalar and vector potentials $V({\bf r},t)$ and ${\bf A}({\bf 
r},t)$ in the single-electron Hamiltonian 
\begin{displaymath}
  H_0({\bf r},t) = \frac{1}{2m}\left[\frac{\hbar}{i}{\bf\nabla}-e{\bf A}({\bf 
  r},t)\right]^2+V({\bf r},t)-\mu
\end{displaymath}
via the matrix equation
\begin{equation} \label{TdBdGE}
i\hbar \frac{\partial  }{\partial t}{u_{n}({\bf r},t) \choose v_{n}({\bf r},t)}  =
\check{\cal{H}}({\bf r},t){u_{n}({\bf r},t) \choose v_{n}({\bf r},t)}.
\end{equation}
Here, the matrix Hamiltonian $\check{\cal{H}}({\bf r},t)$ has $H_0({\bf 
r},t)$ and $-H_0^{*}({\bf r},t)$ in the diagonal, and the pair potential
$\Delta ({\bf r},t)$ and its complex-conjugate in the off-diagonal. The
chemical potential $\mu$ in $H_0$ is that of the reservoir. We neglect all
influences of entropy production associated with current flow on the
chemical potential, because the number of degrees of freedom of the
reservoir is assumed to be much larger than that of the normal leads and the 
superconductor.
Thus, $\mu$ is constant in space and time \cite{kue99}.

AS is a many-body process. For its analysis it is convenient to consider a
non-equilibrium configuration $|T_{l\sigma}\rangle$ of the many-body system
where one quasiparticle  state $(l\sigma)$, characterized by a tripel
$l$ of quantum numbers and spin $\sigma$, is definitely occupied and all
other q.p. states $(n\sigma)$ are occupied according to the equilibrium
Fermi distribution function $f_n$. All interactions that might affect the
spin are neglected. Then, it has been shown \cite{kue69,kue99} with the
help of the TdBdGE (\ref{TdBdGE}) that the expectation values $\langle
T_{l\sigma}|\rho({\bf r},t)|T_{l\sigma}\rangle$ and $\langle
T_{l\sigma}|{\bf j}({\bf r},t)|T_{l\sigma}\rangle$ of the
many-body charge- and current-density operators satisfy the relation
\begin{displaymath}
  \frac{\partial}{\partial t}\langle T_{l\sigma}|\rho({\bf
r},t)|T_{l\sigma}\rangle + {\rm div}\langle
T_{l\sigma}|{\bf j}({\bf r},t)|T_{l\sigma}\rangle
\end{displaymath}
\begin{displaymath}
  = -4\frac{e}{\hbar }{\rm Im}\left[\Delta^*({\bf r},t)u_l({\bf r},t)v_l^*({\bf 
r},t)\right](1 - f_l) + {\rm div}{\bf j}_{sl}.
\end{displaymath}
The electron and hole
wavefunctions $u_l$ and $v_l$ satisfy eq.\ (\ref{TdBdGE}), and
${\bf j}_{sl}$ is the supercurrent density induced by the momentum
transfer from the q.p.\ in $(l\sigma)$ to the superconducting condensate by
AS. The requirement that charge is conserved in the many-body system
results in the source equation for the supercurrent density
\begin{equation} \label{qpsc}
 {\rm div}{\bf j}_{sl}
= 4\frac{e}{\hbar }{\rm Im}[\Delta^*({\bf r},t)u_l({\bf r},t)v_l^*({\bf r},t)]
(1 - f_l).
\end{equation}
A shifted Fermi sphere (or its equivalent in Q2D and Q1D conductors)
represents the current-carrying many-body configurations in the two parts
of the normal current leads that are parallel to the $z$-axis and connected
to the superconductor. (We do not consider the parts of the normal current
leads bent towards the reservoir.) In this
non-equilibrium distribution of electrons
above the Fermi surface in states with positive momentum in $z$-direction
and unoccupied states with negative $z$-momentum below the Fermi surface
the force of the electric field from the battery is balanced by the
frictional forces from electron-phonon interaction.
In the following we try to obtain the
details of normal current $\longleftrightarrow$ supercurrent conversion by
studying the motion of electron and hole wavepackets that are part of the
shifted Fermi sphere. The quasiparticles in this normal-state configuration
are uncorrelated. Thus, the total current in the closed circuit is the sum
of the currents from the individual wavepackets.

The $z$-momenta of the electrons (+) and holes ($-$) in the shifted Fermi 
sphere are
$\hbar k^{\pm}$, with $k^{\pm}(E) = [k_{zF}^2 \pm E2m/\hbar^2]^{1/2}$.
Here, $k_{zF} = [k_F^2-(n_x\pi/L_x)^2-(n_y\pi/L_y)^2]^{1/2}$ is the
$z$-component of the Fermi wavenumber $k_F \equiv [2m\mu/\hbar^2]^{1/2}$,
and $(n_x\pi/L_x)$ and $(n_y\pi/L_y)$, $n_{x,y}$ integers, are the
wavenumbers of the standing waves between the rigid walls that limit the
metals in $x$- and $y$-directions. The energy $E$ of both electrons and holes
is positive and measured relative to
the surface of the unshifted Fermi sphere at the chemical potential $\mu$.
For normal current densities below the critical current densities
of conventional superconductors all $E$ are less than the modulus $\Delta$
of the pair potential $\Delta(z) \approx
\Delta\cdot\Theta(z)\Theta(L_z-z)$, where $\Theta(z)$ is the Heavyside
function.

The construction of the representative wavepacket ensemble, whose motion
and conversion into a supercurrent models the elementary process of current
flow in the {\em N/S/N} circuit, starts with a normalized electron
wavepacket, localized around $z_0 < 0$ in the normal current lead to the
left of the superconductor at time $t=0$. In the center of the wavepacket
the energy is $E_l$. We choose a Gaussian spectral function 
\begin{displaymath}
  D(E) = \frac{a_z}{\sqrt{2\pi}}\,e^{-[k^+(E)-k^+(E_l)]^2a_z^2/2}e^{-i[k^+(E)-k^+(E_l)]z_0};
\end{displaymath}
the position-uncertainty parameter $a_z \ll |z_0|$ is chosen so
large that the related energy spread of the wavepacket, $\delta E = \hbar^2
k_{zF}/ma_z$, is less than $\Delta - E_l$. Solutions of eq.\ (\ref{TdBdGE}),
where $V$ and ${\bf A}$ are neglected and $\Delta({\bf r},t)$ is approximated
by the real $\Delta(z)$ (thereby neglecting repercussions of the q.p.-induced
supercurrent on the q.p.\ and on itself, assuming sufficiently small $L_x$ 
and $L_y$)
are multiplied by the spectral
function $D(E)$, integrated over all energies, and matched at the left
$N/S$ interface, i.e. $z=0$, in the usual Andreev approximation, neglecting
terms of the order of $\Delta/\mu$ outside the exponentials. In so doing,
energy-dependent functions are Taylor expanded around $E_l$ up to first
order in $(E-E_l)$. This affects especially $k^{\pm}(E)$ and the amplitude
of the Andreev-reflection probability 
\begin{displaymath}
  \gamma(E) \equiv e^{-i\arccos E/\Delta} \approx \gamma(E_l)e^{(i/\hbar)(E-E_l)\tau_l},
\end{displaymath}
where $\tau_l = \hbar[\Delta^2-E_l^2]^{-1/2}$ is the time for one
electron $\rightarrow$ hole-scattering event and the associated formation of
a Cooper pair, see eqs.\ (\ref{unl},\ref{vnl}). The resulting electron and hole
wavepackets $u_{NL}({\bf r}, t)$ and $v_{NL}({\bf r},t)$ in the left normal 
current lead, $z<0$,
and the exponentially decaying solutions $u_{SL}({\bf r},t)$ and 
$v_{SL}({\bf r},t) $ in $z>0$,
that contribute to the source equation (\ref{qpsc}) essentially in the left
half of the superconductor, turn out to be
\begin{eqnarray}
u_{NL}& = & w_le^{ik_l^+z}e^{-[z_0-z+v_l^+t]^2/2a_z^2},\label{unl}\\
v_{NL}& = & w_l\gamma(E_l)e^{ik_l^-z}
e^{-[z_0+(v_l^+/v_l^-)z+v_l^+(t-\tau_l)]^2/2a_z^2},\label{vnl}\\
u_{SL}& = & w_le^{ik_{zF}z}e^{-\kappa_l z}e^{-[z_0+v_l^+t]^2/2a_z^2},
\label{us} \\
v_{SL}& = & w_l\gamma(E_l)e^{ik_{zF}z}
e^{-\kappa_l z}e^{-[z_0+v_l^+(t-\tau_l)]^2/2a_z^2},\label{vs}
\end{eqnarray}
where
\[
w_l\equiv \frac{2}{(L_xL_ya_z\sqrt{\pi})^{1/2}}\sin\left(\frac{n_x\pi}{L_x}x\right)
\sin\left(\frac{n_y\pi}{L_y}y\right)e^{-iE_lt/\hbar},
\]
$k_l^{\pm}\equiv k^{\pm}(E_l),$ $v_l^{\pm} \equiv \hbar k_l^{\pm}/m$,
and $\kappa_l \equiv (\Delta^2-E_l^2)^{1/2}/\hbar v_{zF} = 1/\tau_l v_{zF}$,
with $v_{zF} \equiv \hbar k_{zF}/m$. [For the sake of simplicity the
complex wavenumbers in $u_{SL}$ and $v_{SL}$ have not been Taylor expanded
in $(E-E_l)$ but rather taken at $E_l$ right away.]

Identifying the wavefunctions $u_{SL}$ and $v_{SL}$ from eqs.\ (\ref{us})
and (\ref{vs}) with the $u_l$ and $v_l$ of the source equation (\ref{qpsc})
and integrating that equation from $z=0$ to $z$ yields the density of the
supercurrent in $z$-direction induced in the left half of the
superconductor by AS of the electron wavepacket $u_{NL}$ into the hole
wavepacket $v_{NL}$ as
\begin{eqnarray} \label{js}
{\bf j}_{sl,L} &=& {\bf e}_z(2ev_{zF})|w_l|^2[1- e^{-2\kappa_l z}]\nonumber\\
&&\times e^{-\{[z_0+v_l^+(t-\tau_l/2)]^2 + (\tau_l v_l^+/2)^2\} /a_z^2}(1-f_l).
\end{eqnarray}
Here we have assumed that $L_z \gg 1/\kappa_l$. In the opposite case one
would have to add a second source term on the r.h.s. of eq.\ (\ref{qpsc})
which would contain the contribution from the solution 
$u_{SR}({\bf r},t), v_{SR}({\bf r},t)$ 
of eq.\ (\ref{TdBdGE}) for $0<z<L_z$ that
matches to the current-carrying q.p.\ wavepackets in the right normal
current lead at $z=L_z$ and decays exponentially with increasing distance
from $L_z$. This solution determines the supercurrent in the
right half of the superconductor essentially, and its energy and
amplitude are obtained as $E_l$ and $w_l$ from
the requirement that the two supercurrent densities, computed from
$u_{SL}v_{SL}^*$ and $u_{SR}v_{SR}^*$, join smoothly at all times somewhere
within the superconductor. (Since the many-body configurations in the left
and right normal current leads are uncorrelated, only the current
densities, not the wavefunctions, must join smoothly.) Because of the
symmetry of the problem the matching point turns out to be $L_z/2$, and we
obtain
\begin{eqnarray}
u_{SR}& = & w_le^{ik_{zF}z}e^{-\kappa_l(L_z-z)}
e^{-[z_0+v_l^+(t-\tau_l)]^2/2a_z^2},\label{ur} \\
v_{SR}& = & w_l\gamma(E_l)^{-1}e^{ik_{zF}z}e^{-\kappa_l(L_z-z)}
e^{-[z_0+v_l^+t]^2/2a_z^2}. \label{vr}
\end{eqnarray}
The supercurrent density ${\bf j}_{sl,R}$
that results from integrating eq.\ (\ref{qpsc}) from $L_z$ to $z$, with 
$u_{SR}v_{SR}^*$ in the place of
$u_lv_l^*$, has the same form as ${\bf j}_{sl,L}$ of eq.\ (\ref{js}) except
that
$\exp[-2\kappa_l z]$ is being replaced by $\exp[-2\kappa_l(L_z-z)]$. Finally, the
wavepacket solutions $u_{NR}({\bf r},t), v_{NR}({\bf r},t)$ of 
eq.\ (\ref{TdBdGE}) in the right
normal current lead, $z>L_z$, that match to the $u_{SR}({\bf r},t), 
v_{SR}({\bf r},t)$ at the
right interface in $z=L_z$, become
\begin{eqnarray}
u_{NR}& = & w_le^{ik_{zF}L_z}e^{ik_l^+(z-L_z)}
\nonumber\\&&\times 
e^{-[z_0+L_z-z+v_l^+(t-\tau_l)]^2/2a_z^2},\\
v_{NR}& = & w_l\gamma(E_l)^{-1}e^{ik_{zF}L_z}e^{ik_l^-(z-L_z)}
\nonumber\\&&\times 
e^{-[z_0+(v_l^+/v_l^-)(z-L_z)+v_l^+t]^2/2a_z^2}.
\end{eqnarray}

Comparison of the $u_{NL}, v_{NL}$ with the $u_{NR}, v_{NR}$ shows that the
center of the electron wavepacket $u_{NL}$, propagating to the right with
velocity $v_l^+$ in the left normal current lead, and the center of the hole
wavepacket
$v_{NR}$, propagating to the left with velocity $v_l^-$ in the right normal
current lead, hit the left and right interfaces at $z=0$ and $z=L_z$ at the
same time $t_0 = -z_0/v_l^+$, while the hole wavepacket $v_{NL}$,
propagating to the left in $z<0$ with $v_l^-$, and the electron wavepacket
$u_{NR}$, propagating to the right in $z>L_z$, are retarded by the time
$\tau_l$ with respect to the incident wavepackets. This means: electron
$\rightarrow$ hole scattering in the left interface combines the electron
with spin up above the Fermi surface with another spin-down electron below
the Fermi surface to form a Cooper pair at the chemical potential, leaving
the hole with spin up in the left normal region, transferring the momentum
$2\hbar k_{zF}$
to the condensate, and simultaneously in the right
interface a Cooper pair decays
by hole $\rightarrow$ electron scattering into the incident hole and the
reflected electron of total momentum
$2\hbar k_{zF}$.
The supercurrent density ${\bf j}_{sl}$ spreads instantaneously in the
superconductor. Its spatial maximum is at $z=L_z/2$, and its maximum in
time occurs at $t = t_0 + \tau_l/2$. From the equations for ${\bf j}_{sl}$,
$u_{SL,R}$ and $v_{SL,R}$ one sees that the quasiparticle current density
\begin{displaymath}
  {\bf j}_{QPl} \equiv \frac{e}{m}{\rm Re}\left[u_l^*\frac{\hbar}{i}{\bf \nabla}u_l 
- v_l\frac{\hbar}{i}{\bf \nabla}v_l^*\right](1-f_l)
\end{displaymath}
changes into the supercurrent
density
${\bf j}_{sl}$
and vice versa within a distance $1/\kappa_l = v_{zF}\tau_l$ from the
interfaces. Figure 1 shows stills from a computer movie of the described
Andreev-scattering and 
wavepacket-current $\longleftrightarrow$ supercurrent--conversion processes.

By summing the current contributions from the steady flow of  all 
quasiparticle wavepackets
$(l\sigma)$, $l\equiv(E_l,n_x,n_y)$, that corresponds  to the shifted Fermi 
sphere in the normal current
leads, one obtains the total normal current and the total
supercurrent  within the closed $N/S/N$ circuit.

The supercurrent, carried by the condensate in the $S$ layer, involves only 
states
with $|E_n| \geq \Delta$. It continues the current from the electrons and 
holes with energies $E_l < \Delta$ in the normal current leads.
If the total current density exceeds its critical value, i.e.\ if the center of
the Fermi sphere in the normal current leads is shifted by more than 
$\hbar q_{cS}= \Delta m/\hbar k_F$, depairing sets in, and when 
superconductivity has broken down 
the uncorrelated normal-state configuration reigns everywhere in the circuit.
If the single $S$ layer is replaced by a mesoscopic $SNS$ junction, the
many-body configuration in the central $N$ layer is a phase-coherent one
and thus different from the uncorrelated configurations in the normal
current leads. In an $N/SNS/N$ circuit the $SNS$ junction acts as a gapless
superconductor \cite{ish70}. It can carry a dissipation-free Josephson
current through the central $N$ layer via phasecoherent q.p.\ states with 
$|E|<\Delta$ and $|E|
 \geq \Delta$ \cite{gun94}. This current converts as a whole into the total
supercurrent of  the $S$ layers, and vice versa, whereas, according to 
eq.\ (\ref{qpsc}), each uncorrelated q.p.\  from the normal
current leads individually induces its proper fraction of the total 
supercurrent. If the total current density exceeds the critical 
Josephson-current density at a Fermi-sphere shift of
$\hbar q_{cSNS}\approx\hbar/d$, where $d$ is the length of the central 
$N$ layer \cite{bar72},
a voltage drop appears across the $SNS$ junction. 
There are different
models \cite{oct83,kue90,gun94a,ave95,jac99} for $SNS$ junctions with voltage 
drops. They differ with respect to the  implicit
assumptions about the rate and energy range of q.p.\ creation in the central
$N$ layer by supercurrent $\longrightarrow$ quasiparticle-current
conversion. The question of how this rate and range depend upon the
weakening of phase coherence in the $SNS$ junction by energy exchange between
quasiparticles and electric fields, phonons, and thermal
fluctuations like Nyquist-Johnson noise  is presently
investigated.\\

\newpage

\begin{figure}[htb]
\PostScript{0}{8.8}{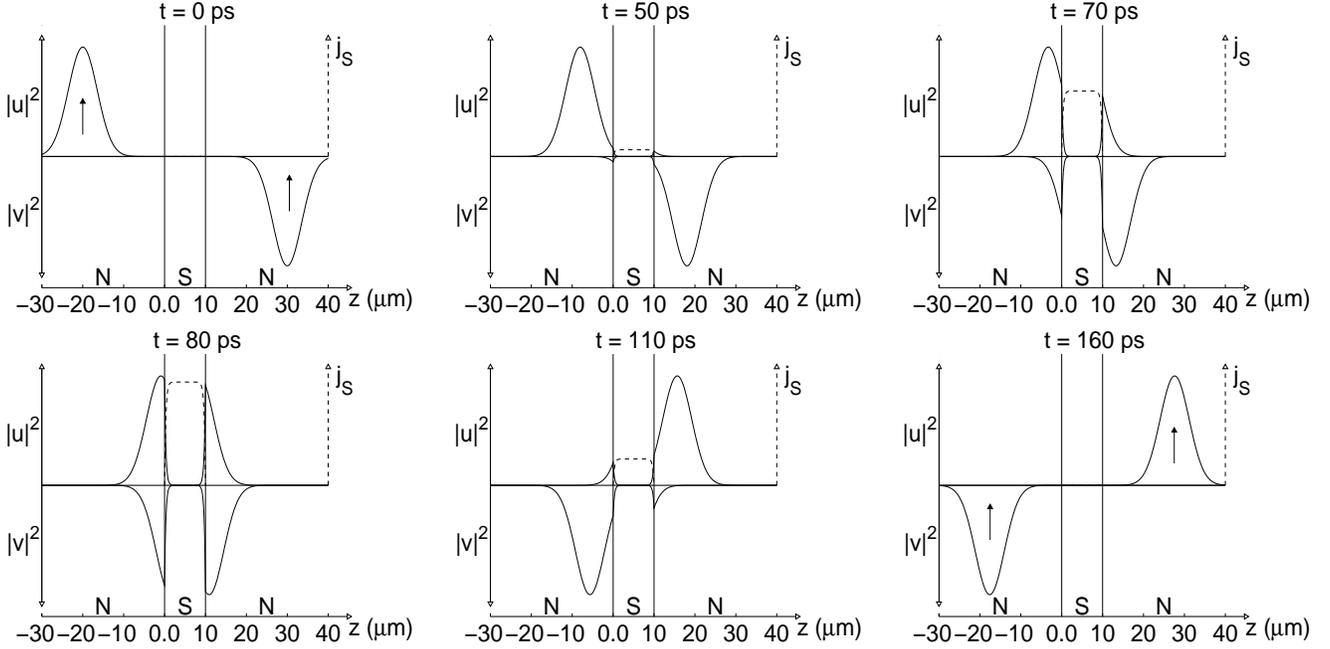}
\caption{Propagation and Andreev scattering of the probability densities $|u|^2$ 
and $|v|^2$ of spin-up electron and hole wavepackets (solid lines) and the induced 
supercurrrent density ${\bf j}_s$ (dashed line). The energy in the center of the 
wavepackets is  $E_l = \Delta/2 = 0.15 \mbox{ meV}$, the spatial wavepacket spread 
is $a_z = 5 \mbox{ $\mu$m}$  and $k_{zF} = 0.9k_F = 2.06 \mbox{ nm}^{-1}$. The time 
elapsed between the 
first and the last picture is 160 ps. The time elapsed between the center of 
the incident left electron (right hole) wavepacket hitting the interface at 
$z=0, (z=L_z=10 \mbox{ $\mu$m})$ and the center of the reflected left hole (right 
electron) wavepacket leaving the superconductor is $\tau_l =
\hbar/[\Delta^2-E_l^2]^{1/2} = 2.53 \mbox{ ps}$.
Note that current flow in a closed $N/S/N$ circuit always involves electron 
$\rightarrow$ hole scattering in one $N/S$ interface and hole $\rightarrow$ electron 
scattering in the other interface. The Andreev-reflected wavepackets may be 
considered as supercurrent-transmitted wavepackets as well.
More stills from other computer movies on electron $\rightarrow$
hole and electron $\rightarrow$ electron scattering for various $E_l$ and
$k_{zF}$ in one interface between semi-infinite $N$ and $S$ layers 
can be viewed in the Internet under 
{\em http://theorie.physik.uni-wuerzburg.de/TP1/kuemmel/ profile.html}.}
\label{label1}
\end{figure}


\end{document}